\documentclass[preprint]{revtex4}
\usepackage{amsfonts,amssymb,epsfig}
\begin{document}
\sloppy
\newcommand{\fdag}{\not\kern-.25em}

\title{
Nucleon-Nucleon Optical Potentials \break and Fusion of
$\pi$N, KN, $\pi\pi$ and NN Systems
\footnote{To appear in the Procedings of the International Workshop on
Resonances in Few-Body Systems,  4-8 September 2000, Sarospatak Hungary,
Few-Body Systems, Supplement, Springer Verlag }
}
\author{ von Geramb H.V.}\email{geramb@uni-hamburg.de}
\author{ Funk A.}
\author{ Faltenbacher A.}

\affiliation{
Theoretische Kernphysik, Universit\"at Hamburg,
Luruper Chaussee 149, D-22761 Hamburg, Germany
}

\begin{abstract}
Several boson exchange potentials, describing the NN interaction $T_\ell\le 300$ 
MeV with high quality, are extended in their  range of applicability 
as NN optical models  with complex local or separable  potentials in r-space 
or as complex  boundary condition models.
We determine in this work  the separable potential strengths or boundary
conditions on the background of the  Paris, Nijmegen-I, Nijmegen-II, Reid93, 
AV18 and inversion potentials. Other hadronic systems, $\pi$N, KN and $\pi\pi$, 
are studied with the same token. We use the latest phase shift  
analyzes SP00, SM00 and FA00 by Arndt {\em et al.} as input and thus extent
the mentioned potential models from 300 MeV to 3 GeV .
The imaginary parts of the interaction account for loss of flux into direct 
or resonant production processes. For a study of  resonances and absorption
the  partial waves wave functions 
with physical boundary conditions are calculated. We display the
energy and radial dependences of  flux losses  and radial probabilities. 
The results lend quantitative  support for the established
mental image of intermediate elementary particle formation in the 
spirit of fusion.
\end{abstract}

\maketitle

\section{Introduction}
 
From the recently fitted high quality  boson exchange models 
their is none which fits the latest NN phase shift analyzes,  which today extents 
to 3 GeV\cite{Arn00}. Several theoretical attempts, which explicitly included 
the Delta and other nucleonic resonances, remained qualitative
and do not meet the required high standard of reproduction of phase shifts or
on-shell t-matrices, as  is needed in their use  for nucleon-nucleus reaction studies.
Potentials  are still very convenient to extent the two nucleon on-shell
t-matrix into the off-shell domain which enters in few and many body reaction
calculations. It is now well established that all the high quality
NN potentials have a very similar off-shell continuation which is practically
the same as implied by local inversion potentials in r-space. 
Repeatedly occur claims, regarding the NN t-matrix off-shell, of sensitivity 
in few body reactions. Practically, these claims end mostly  in 
clear failures of precise on-shell reproductions of t-matrices. We followed
such studies in pp bremstrahlung and microscopic optical model calculations at
low and medium energy and found unanimous support for need of a precise
on-shell reproduction with little effects from  far off-shell t-matrices. 
This is what our approach satisfies.

Furthermore, the $\pi$N boson exchange models and older 
phenomenological separable potentials  
describe the Delta resonance rather as a molecular system,
composed of a well separated $\pi$ and nucleon, and not as a compactly fused 
elementary particle. Contrary to these findings
support elastic scattering inversion potentials in the [3/2,3/2]  
channel the intermediate formation
of a fused $\pi$N  system,  the Delta resonance. 

Since,  already at low energy, production and in-elasticity
is important for many elementary systems 
we develop optical potentials for NN, $\pi$N, $\pi\pi$ and  KN  scattering. 
The details of how these
potentials are calculated with a stringent link to {\em  experimental phase shifts} 
is contained in Sect. 2. The  complex optical model potentials are then used to
calculate wave functions, radial probabilities
and the radial losses of flux from the continuity equation.

What we expect from these studies is the support of 
existing mental images of formation and fusion  of elementary particles.
The transition into a combined object, viewed in terms of the relative distance
between the centers of particles, is to be signaled by a concentration of loss of
flux from the entrance channel and thus disappearance into the QCD sector and
deeper down into complex particle production processes. We expect a
typical distance of  $0.3-0.6$ fm with the implication that 
the charge form factors overlap significantly, 80-95\%. 
Ad hoc, this picture is not supported for $\pi$N boson exchange potentials, 
at least in studies of the $\Delta$ resonance. 
To illustrate these findings we show  samples  in Sect.3.    

The elastic partial wave NN phase shifts are now available from Arndt {\em et al.}
for energies $<3$ GeV. Caught from  the spirit of inverse scattering, we generate a
quantitative optical model for any of these NN partial waves 
with a versatile program which allows studies of
complex separable potentials on the background  of the
existing NN potentials like Paris, Nijmegen and our own inversions.
The described  procedure,  Sect 2, is  the inverse scattering branch of
soft obstacle identification with elastic scattering data and their
partial wave phase shift decomposition respectively. With the experimentally
determined phase shifts and thus the on-shell t-matrix,
an  inversion algorithm is a mean to continue these on-shell values into the
off-shell domain within  classes of potentials. The class is limited in the
present calculations to be a range of well established, real valued in r-space,
OBEP (Paris, Nijmegen-II, Nijmegen-I, Reid-93, AV18, ESC-96) 
or inversion reference potentials  and  
separable potentials  with complex strengths.  The separable potential strengths
are determined with an inversion algorithm which guarantees  full
agreement with any experimental phase shift in  any partial wave and energy.
A versatile range of options for the separable potential form factors 
permit also studies of boundary condition models which suit 
distinctions between  hadronic and QCD domains.
      
\section{The Optical Model Potentials}

NN scattering is formally described by the Bethe-Salpeter equation
\begin{equation} \label{BeSal}
{\cal M} = {\cal V} + {\cal V}{\cal G}{\cal M}\ .
\end{equation}
It serves as  ansatz for
the Blankenbecler-Sugar reduction which is obtained from the
integral equation  (\ref{BeSal})
in terms of four-momenta
\begin{equation}
{\cal M} ( q^\prime,q;P ) = {\cal V} ( q^\prime,q;P ) + 
\int d^4k\ {\cal V} ( q^\prime,k;P )\  {\cal G} (k;P)\ {\cal M} ( k,q;P)\ 
\end{equation}
using the propagator
\begin{equation}
{\cal G} (k;P) = { i \over (2 \pi)^4} \left[ { \frac12\fdag P
+ \fdag k + M \over (\frac12 P + k)^2 - M^2 + i \varepsilon}\right]^{(1)}
\left[ { \frac12\fdag P
+ \fdag k + M \over (\frac12 P + k)^2 - M^2 + i \varepsilon}\right]^{(2)}\ .
\end{equation}
The superscripts refer to the nucleon (1) and (2) respectively, and in 
the CM system, $P = (\sqrt{s},0)$,  with  total energy $\sqrt{s}$. 
The reduction of the propagator $\cal G$
uses the covariant form
\begin{equation}
{\cal G}_{\rm BbS} (k,s) = - {\delta (k_0) \over (2 \pi)^3 } { M^2 \over E_k}
{\Lambda_+^{(1)} ({\bf k}) \Lambda_+^{(2)} (-{\bf k}) \over
\frac14s - E_k^2 + i \varepsilon}\ ,
\end{equation}
where only the positive energy projector is used.
Finally, a three-dimensional equation is obtained from
\begin{equation}
{\cal M} ({\bf q}^\prime,{\bf q}) = {\cal V} ({\bf q}^\prime,{\bf q}) +
\int { d^3k \over (2 \pi)^3 }{\cal V} ({\bf q}^\prime,{\bf k})
 { M^2 \over E_k} {\Lambda_+^{(1)} ({\bf k}) \Lambda_+^{(2)} (-{\bf k}) \over
{\bf q}^2 - {\bf k}^2 + i \varepsilon}
{\cal M} ({\bf k},{\bf q})\ .
\end{equation}
Taking matrix elements with 
only positive energy spinors yields  
the  minimum relativity form
\begin{equation}
{\cal T} ({\bf q}^\prime,{\bf q}) =  {\cal V} ({\bf q}^\prime,{\bf q}) +
\int { d^3k \over (2 \pi)^3}  {\cal V} ({\bf q}^\prime,{\bf k})
 { M^2 \over E_k} {1 \over {\bf q}^2 - {\bf k}^2 + i \varepsilon}
{\cal T} ({\bf k},{\bf q})\ .
\end{equation}
With the substitutions
\begin{equation}
 T ({\bf q}^\prime,{\bf q})
= \left( M \over E_{q^\prime} \right)^{\frac12} 
{\cal T} ({\bf q}^\prime,{\bf q})
\left( M \over E_{q} \right)^{\frac12}
,\ 
V ({\bf q}^\prime,{\bf q})
= \left( M \over E_{q^\prime} \right)^{\frac12}
{\cal V} ({\bf q}^\prime,{\bf q})
\left( M \over E_{q} \right)^{\frac12},
\end{equation}
we obtain a  Lippmann-Schwinger equation for the t-matrix 
\begin{equation}
 T ({\bf q}^\prime,{\bf q}) =  V ({\bf q}^\prime,{\bf q}) +
\int { d^3k \over (2 \pi)^3}  V ({\bf q}^\prime,{\bf k})
 {M \over {\bf q}^2 - {\bf k}^2 + i \varepsilon}
 T ({\bf k},{\bf q})\ .
\end{equation}
The relation between t-matrix and potential, or free wave function
and scattering wave function, $T\Phi=V\Psi$  
gives the  Schr\"odinger wave equation 
\begin{equation} \label{relschroe}
\left[-\Delta +M\, V({\bf r})-k^2\right] \psi({\bf r},  k)=0\ ,
\end{equation}
where  the reduced mass 
\begin{equation}
M={2\mu\over \hbar^2}={2\over \hbar^2}{m_1m_2\over m_1+m_2}\ .
\end{equation}
must be used for consistency with reference potentials. 
A careful and consistent treatment of  $(M/E)$ factors in the transformation 
are important for  use of t-matrices in few and many  body calculations
where relativity is of importance.
Minimal relativity enters only in the calculation of $k^2$,
\begin{equation}
M_{12}^2=(m_1+m_2)^2+2m_2T_{\mbox{Lab}}=
\left(\sqrt{k^2+m_1^2}+ \sqrt{k^2+m_2^2}\right)^2\ ,
\end{equation}
while the relative momentum in CM system is
\begin{equation}
k^2 = { \displaystyle m_2^2 ( T_{Lab}^2 + 2 m_1 T_{Lab}) \over
\displaystyle ( m_1 + m_2 )^2 + 2 m_2 T_{Lab}}\ ,
\end{equation} 
which, for equal masses, reduces to $ k^2 =  s/4- m^2$.

\subsection{Algorithm for  Separable Optical and Boundary Condition Models}

We distinguish three Hamiltonians: 
i.) A reference  Hamiltonian $H_0:=T+V_0$  
with elastic scattering solutions $\psi_0:=\psi_0^+(\vec r,\vec 
k,E)$, with real phase shifts (equivalent to
a unitary S-matrix in the elastic channel),
of which any quantity, such as wave functions and phase shifts, is readily 
computed;
ii.) A projected  Hamiltonian
$H_{PP}:=P H P$,  with  $P+Q=1$ and $Q:=\sum_{i=1}^N |\phi_i><\phi_i|$.
The Q-space functions $|\phi_i(r)>$ are doorway states
describing the  QCD entrance sector  
from finite size nucleons, mesons and possible other particles.
Meson creation and annihilation occur only in Q-space and deeper down.
iii.) Full optical model Hamiltonian 
${\cal H}:=T+V_0+{\cal V}(r,r';lsj,E)$, with scattering solutions 
$\Psi^+:=\Psi^+(\vec r,\vec k,E)$.
They shall match asymptotically  continuous energy fit solution of 
elastic channel S-matrices. In the present case, the  experimental 
complex phase shifts are taken from SAID\cite{SAID}.
The reference potential $V_0$ and ${\cal V}(r,r';lsj,E)$ are to be specified
in  detail with any of the applications.

\subsection{Towards Full Potential Model}

Using the scattering equations, $(E-H_0)\psi_0=0$ and
$(E-H_{PP})\psi_P=0$
\begin{equation}
(E-H_{PP}-H_{QP}-H_{PQ}-H_{QQ})
\psi_P=-H_{QP}\psi_P
\end{equation}
or as Lippmann-Schwinger equation, 
\begin{eqnarray}
\psi_P&=&\psi_0-{1\over(E^+-H_0)}H_{QP}\psi_P
\nonumber \\ &=&\psi_0-\sum_jG^+|j><j|H_{QP}\psi_P.
\end{eqnarray}
Orthogonality $PQ=QP=0$ yields 
\begin{equation}
<i|\psi_P>=0=<i|\psi_0>-<i|G^+ H_{QP}|\psi_P>,
\end{equation}
and 
\begin{equation}
<j|H_{QP}|\psi_P>=\sum_i^N \{ <i|G^+|j>\}^{-1}<i|\psi_0>.
\end{equation}
The P-space solutions, in terms of $\psi_0$, are
\begin{equation}
\psi_P=\psi_0-\sum_{ij}^N G^+|i>{1\over <j|G^+|i>}<j|\psi_0
\end{equation}
with  a separable potential
\begin{equation}
|i>{1\over <j|G^+|i>}<j|=|i>\lambda_{ij}<j|=:\Lambda_{ij}(r,r')
\end{equation}
and
\begin{equation}
\psi_P=\psi_0-\sum_{ij}^N G^+ \Lambda_{ij}\psi_0
\label{eqn1}
\end{equation}
The Q-space definition  gives  a unique value for the
separable potential $\Lambda_{ij}(r,r')$ and 
its strengths $\lambda_{ij}(lsj,E)$.
The result looks like a first order Born approximation but is exact.
However, in the next step we allow the strength matrix 
\begin{equation}
\lambda_{ij}=\{<n|G^+|m>\}^{-1}_{ij}
\end{equation}
to vary  with the implication  to
reproduce  asymptotically the experimental phase shifts of the full Hamiltonian 
\begin{equation}
H_{PP}\rightarrow {\cal H}.
\end{equation}
In case of non unitary S-matrices a complex optical model is the result of
matching 
\begin{equation}
\Psi_{\cal H}\sim \psi_P\sim {1\over 2i}(-h^- +h^+ S(k)).
\end{equation}
From 
\begin{equation}
\frac 1{2i}{h^+(kR)(S(k)-S_0(k))= 
\sum_{ij}G^+|\phi_i>\lambda_{ij}<\phi_j|\psi_0^+>}
\end{equation}
we obtain the strengths $\lambda_{ij}$. Next,
the separable potential is transformed into    
\begin{equation} {\cal V}(r,r'):=\Lambda{1\over(1-G^+\Lambda)}
\end{equation}
for which the  Lippmann-Schwinger equation
\begin{equation}
\Psi_{\cal H}=\psi_0+G^+{\cal V}\Psi_{\cal H}.
\end{equation}
holds.

\subsection{Technical Details}

The radial wave functions (we identify the complete channel set of
quantum numbers by $\alpha$)  of the reference potential
\begin{eqnarray} 
u''_\alpha(r,k)=\left({L(L+1)\over r^2}+{2\mu\over \hbar^2 } V_a(r,\alpha)
\right. \nonumber \\ 
-\left.\left({V_b'(r,\alpha)\over 1+2V_b(r,\alpha)}\right)^2-{k^2\over  
1+2V_b(r,\alpha)}\right)
u_\alpha(r,k) 
\end{eqnarray}
are numerically integrated for single and coupled channels by Numerov. The
potentials $V_a,V_b,V_b'$ are available for the quantum numbers (LSJT) for the 
Paris, Nijmegen, Argonne and our own inversion  r-space potentials
Paris, Nij1, ESC96 all $V_b\neq 0$, and  Nij2, Reid93, AV18, HHinv all $V_b=0$.
The physical solutions are matched
asymptotically to Riccati Hankel functions
\begin{equation}
u^+_\alpha(r,k)\sim \frac i2 (h^-_\alpha(r,k)-h^+_\alpha(r,k)S^0_\alpha(k))
\end{equation}
and normalized 
\begin{equation}
\psi_\alpha^+(r,k)={u_\alpha^+(r,k)\over\sqrt{ 1+2V_b(r,\alpha)}}.
\end{equation}
The outgoing wave Jost solution 
\begin{equation}
{\cal J}^+_\alpha(r,k)\sim h^+_\alpha(r,k)
\end{equation}
is calculated by the same token as the physical solution and
they are used in the reference potential Green's function 
\begin{equation}
G^+_\alpha(r,r',k)=\left\{\begin{array}{l}
-\left(2\mu\over \hbar^2 \right)\frac 1k\
\psi^+_\alpha(r,k)\widetilde{\cal J}^+_\alpha(r',k),\  r<r' \\[0.2cm]
-\left(2\mu\over \hbar^2 \right)\frac 1k
{\cal J}^+_\alpha(r,k)\widetilde{\psi}^+_\alpha(r',k),\  r>r'
\end{array}\right. 
\end{equation}
Therewith, at the asymptotic matching radius $r=R$, the strengths
matrix $\lambda_\alpha(k)$ is calculated from  
\begin{eqnarray}  
\Psi_\alpha^+(R,k)=\psi^+_\alpha(R,k)+\\[0.2cm] 
\int_0^\infty G^+_\alpha(R,r',k)|r',\alpha>dr'\lambda_\alpha(k)
\int_0^\infty <r',\alpha|\psi^+_j(r',\alpha)dr'. \nonumber
\end{eqnarray}
Using 
\begin{eqnarray} 
\Psi_\alpha^+(R,k)-\psi^+_\alpha(R,k)=\frac 1{2i}h_\alpha^+(R,k)(S_\alpha
(k)-S^0_\alpha(k)) \\[0.2cm] =
\int_0^\infty G^+_\alpha(R,r)|r,\alpha>\,Sunitarydr \lambda_\alpha(k)\int_0^\infty<r,\alpha
|\psi^+_\alpha(r,k)\;dr.  \nonumber
\end{eqnarray}
and the final separable potential strengths 
\begin{equation}
\sigma_\alpha(E(k))=\left(1-\lambda_\alpha(k)
\int_0^\infty\int_0^\infty<r,\alpha|G^+_\alpha(r,r',k)|r',\alpha >dr\, dr'
\right)^{-1}\lambda_\alpha(k).
\end{equation}
The Lippmann-Schwinger equation for the full problem solutions are generated with
the reference potential solutions and its Green function plus the separable
potential with the strengths $\sigma_\alpha(k)$
\begin{eqnarray}
\Psi_\alpha^+(r,k)=\psi^+_\alpha(r,k)+  \\[0.2cm]
+\int_0^\infty G^+_\alpha
(r,x,k)|x,\alpha>dx\;\sigma_\alpha(E(k))\;
\int_0^\infty <y,\alpha|\Psi^+_j(y,\alpha)\;dy \nonumber
\end{eqnarray}
Several options of separable potential are available. In principle any rank is
admitted but this requires to determine the strengths $\lambda_\alpha(k)$ from data
at several energies around the mean energy $E(k)$. The rank $>1$ 
option can be succesfull for single channel cases but suffers from lack of
energy dependence in data for coupled channels. As we may use any radial
form factor we found rank one separable potentials in any case the better option
and they are: a.)
Harmonic oscillator radial wave functions $\phi_\alpha(r,n=1,\hbar\omega)$;
b.) Normalized Gaussians $N_0\exp -(r-r_0)^2/a_0^2$ with
$r_0=0.3-0.8$ fm, $a_0= 0.1$ fm,
with the option to put the reference potential equal to any value for $r\leq r_0$,
simulates a smooth boundary condition around $r_0$;
c.) A step function $\phi(r_0,\alpha)=1,\ \phi(r\pm h,\alpha)=0.5$ and zero else, is
very suitable to determine the boundary conditions at a chosen  radius $r_0$.
It serves as surface for the transition from the hadronic into the
QCD domains. It is the usual boundary condition model. 
Obviously, these examples of  form factors are may be 
replaced  by microscopic QCD inspired models with a wide range of theoretical
inspirations\cite{Sta00}.

\subsection{Derived Quantities}

The probability density of the full $\cal H$ problem
\begin{equation} 
\rho(r,\alpha,E)  = Trace\ \Psi^\dagger (r,\alpha,E)\Psi(r,\alpha,E)r^{-2}
\end{equation}
and the loss of flux, from the continuity equation
\begin{equation}
{\partial\over \partial t}\rho(r)+(\vec \nabla\cdot\vec j)=0,
\end{equation}
\begin{equation}
(\vec \nabla\cdot\vec j) : = Trace\ {i\over \hbar}
<\Psi(\alpha,E)|{\cal V}(.,r)-{\cal V}^\dagger(r,.)|\Psi(\alpha,E)>.
\end{equation}
are interesting to display. 
\newpage
\section{ Applications for $\pi$N and $\pi\pi$ Systems}

We concentrate in this contribution  upon mesonic channels and refer for the
NN results to the contribution by Andreas Funk.
In Fig. 1 we show a sample of the latest SAID $\pi$N continuous phase shift solutions
(crosses are single energy solutions) and the optical model potentials
reproduce exactly the continuous soltions (full lines)\cite{SAID}.
The rapid changes  with energy  are signaling  resonances which have its
correspondences in a strong build-up and concentration of probability at short
distances $\sim 0.25-0.6$ fm. Similar to 
the probabibilties behaves the loss of flux which is concentrated in the same radial
region,  left and right columns. Best fusion examples are seen in the 
$\ P_{33},\ D_{13}$, a highly inelastic resonance is$\ D_{15}$, and no fusion is
seen in $P_{31}$. Some resonances feed strongly into 
other channels, shown with the loss of flux which is highly 
concenrtrated at the same energy, whereas others are predominant elastic
channel resonances, see$P_{33}$.  Similar results are observed in other channel and in
$\pi\pi$ and KN systems \cite{Fal99}. We conclude from our calculations that a 
significant part of resonances, listed in the Data Tables, do not fall into the
category of fused systems but rather show a strong energy dependence with complex
structures in probilities and absorption, there appearances are more like molecules.  
Complex and fused structures, like dibaryon
resonaces, $\pi\pi$ see Fig. 2, and other hybrid excitations, 
are anticipated for many hadronic systems and 
theoretical and experimental work is in progress\cite{Bar00}.      

\newpage
\begin{figure}[ht]
\centering
\epsfig{file=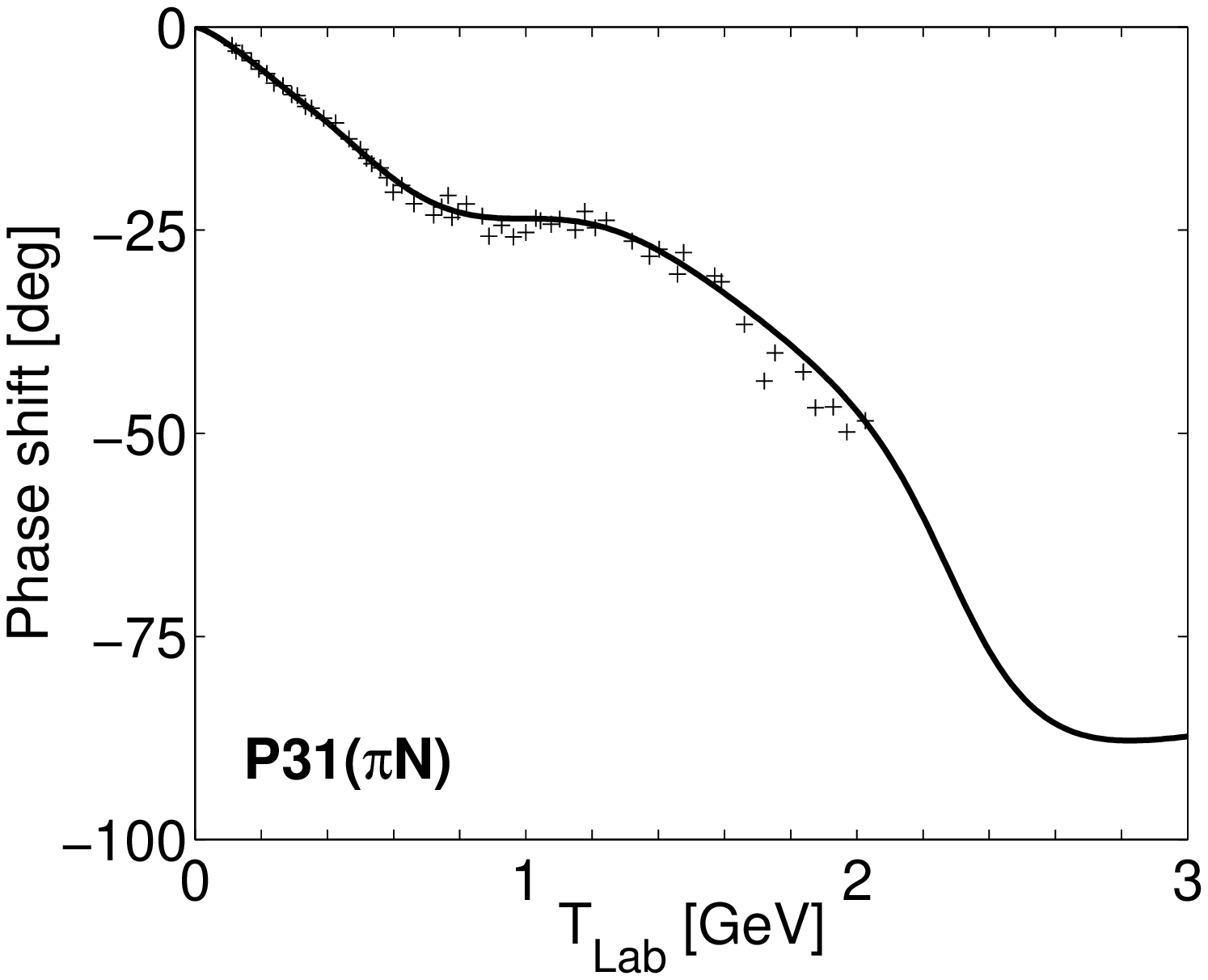,scale=0.37}
\epsfig{file=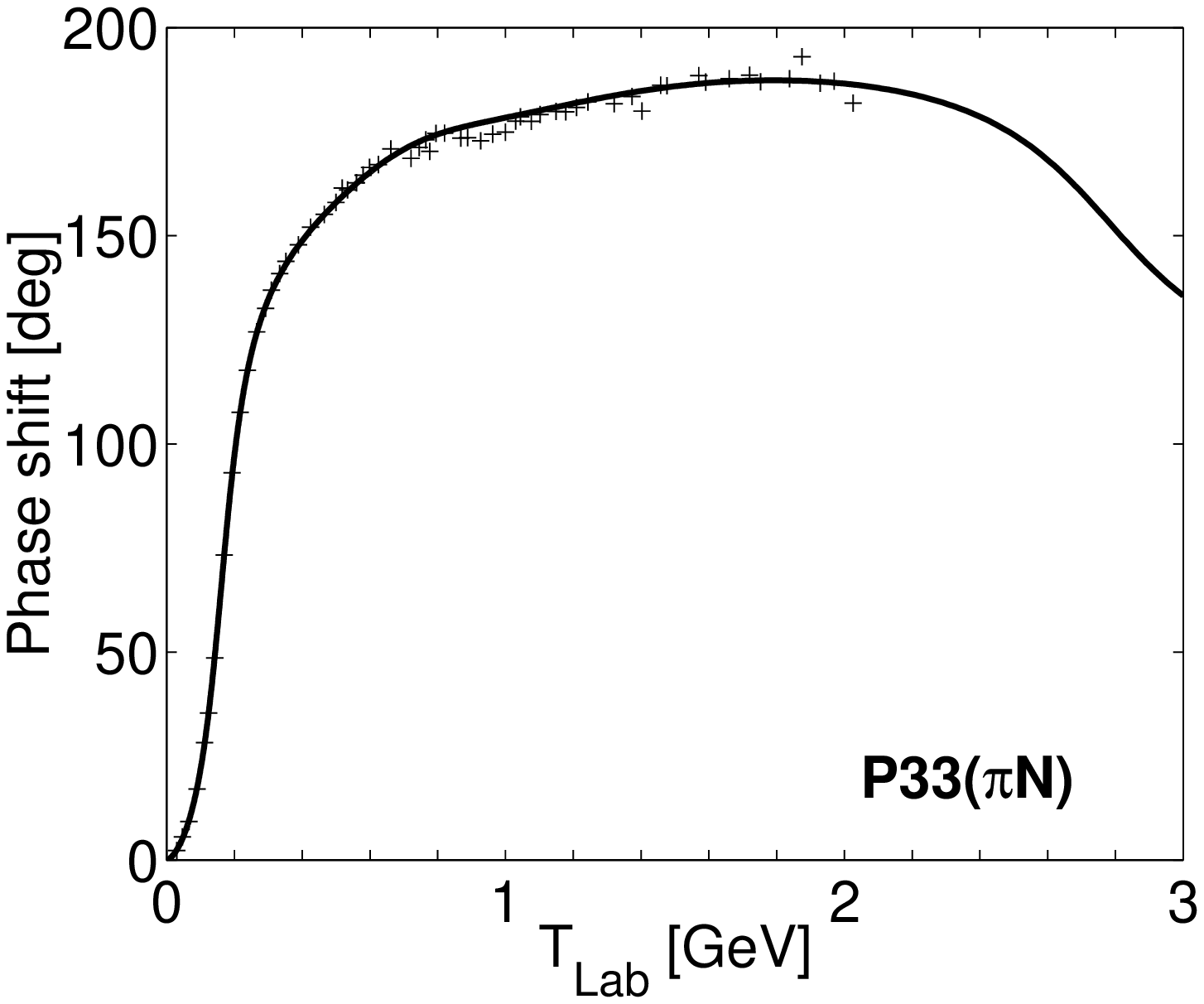,scale=0.37}
\epsfig{file=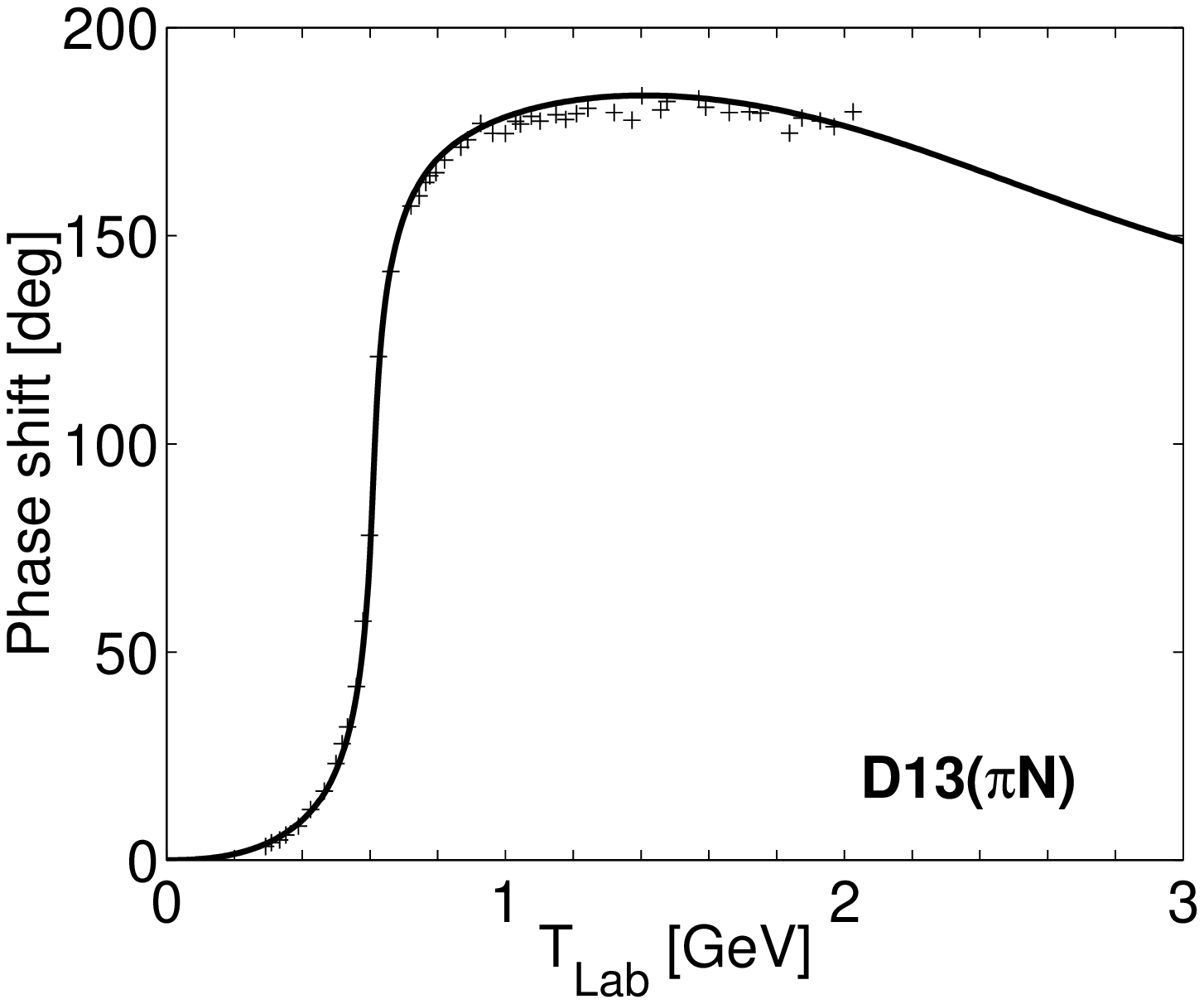,scale=0.37}
\epsfig{file=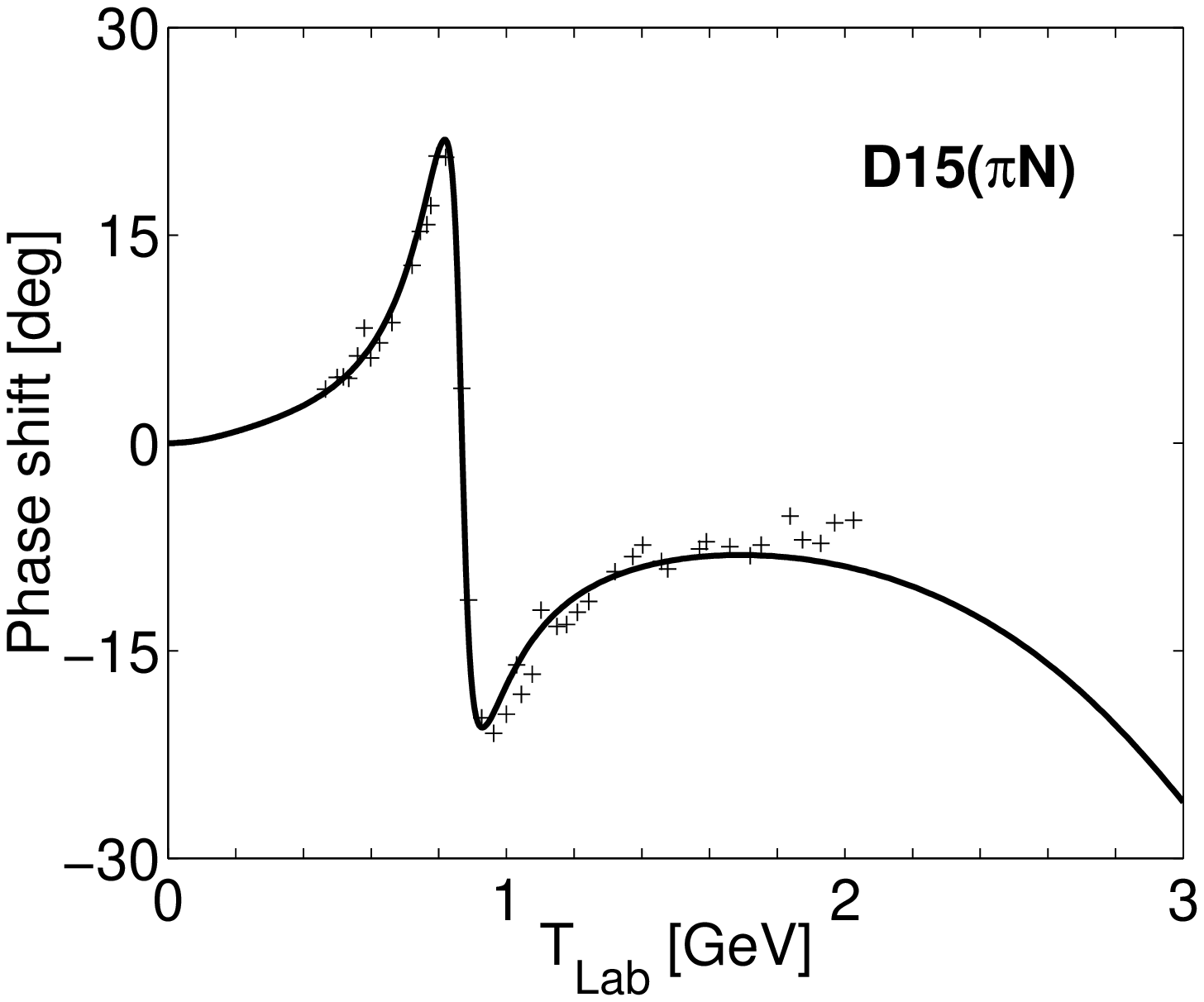,scale=0.37}
\epsfig{file=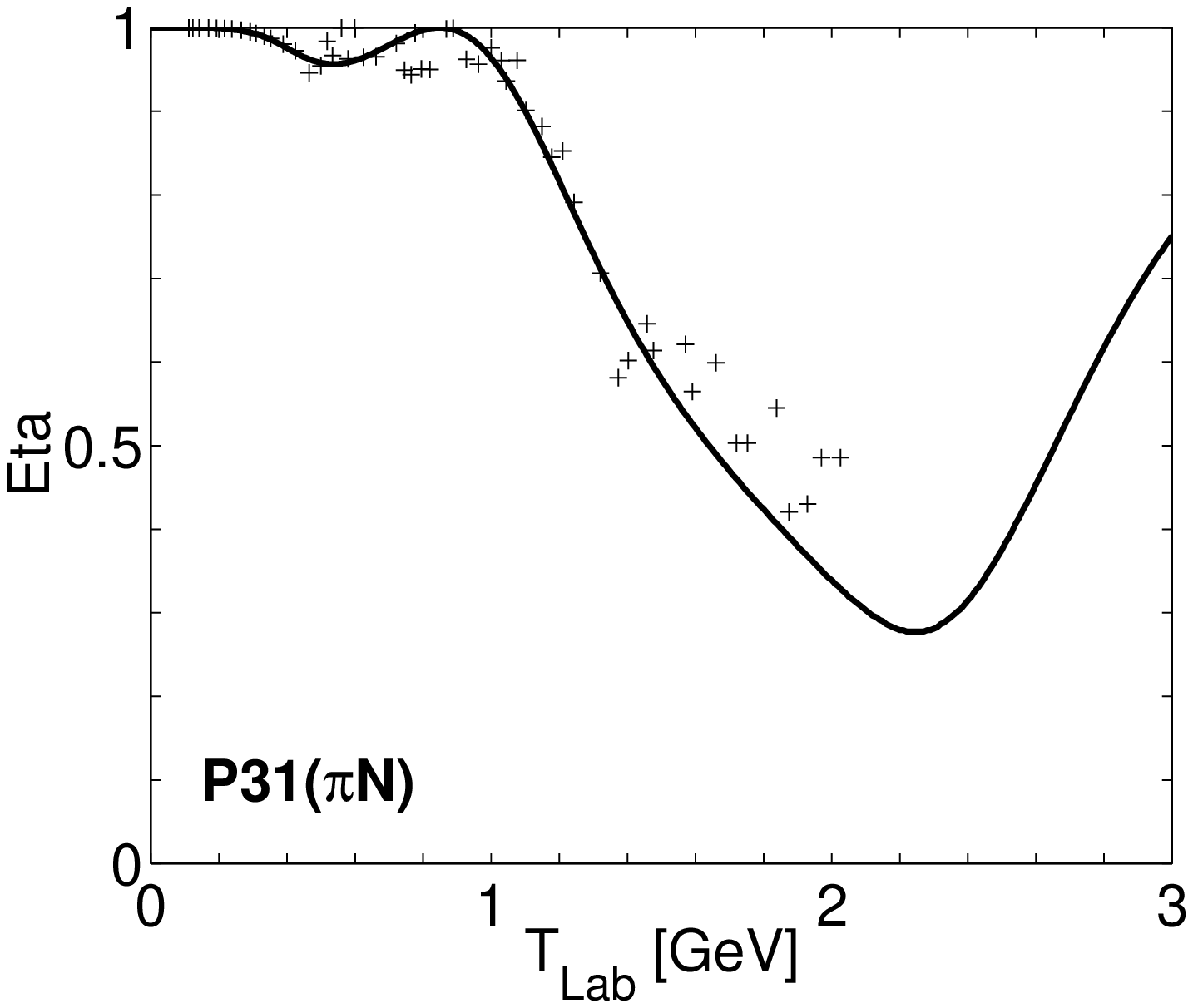,scale=0.37}
\epsfig{file=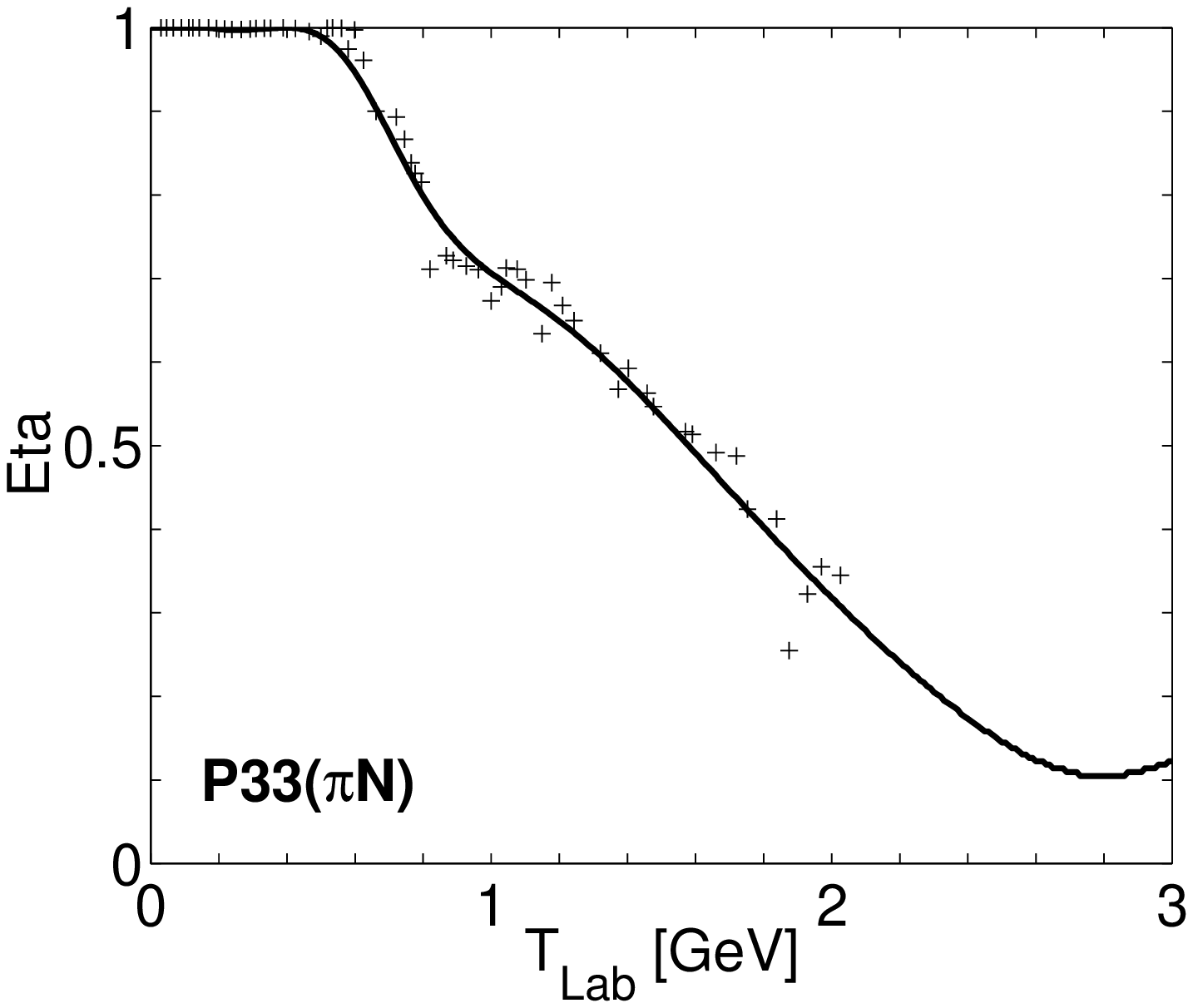,scale=0.37}
\epsfig{file=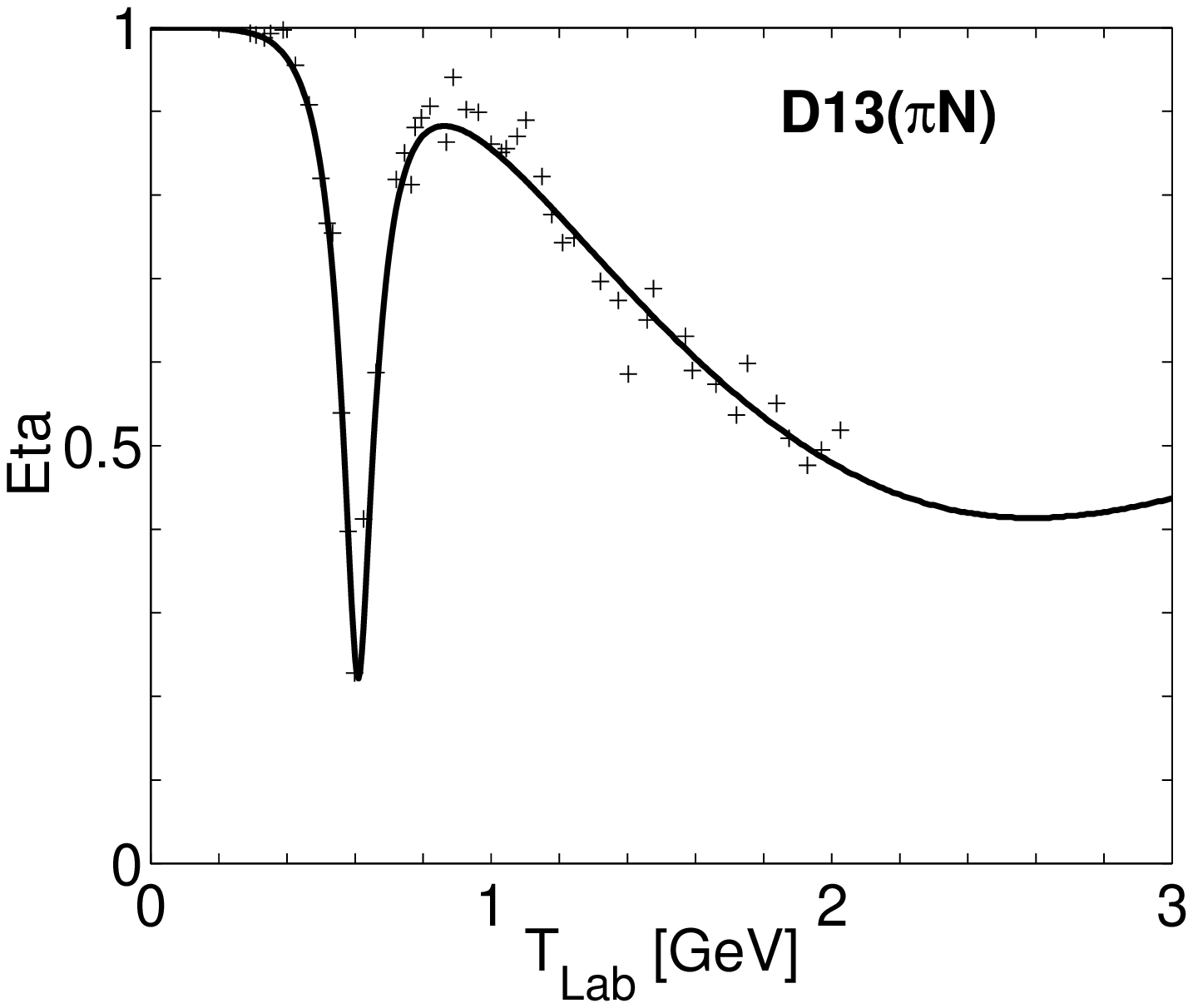,scale=0.37}
\epsfig{file=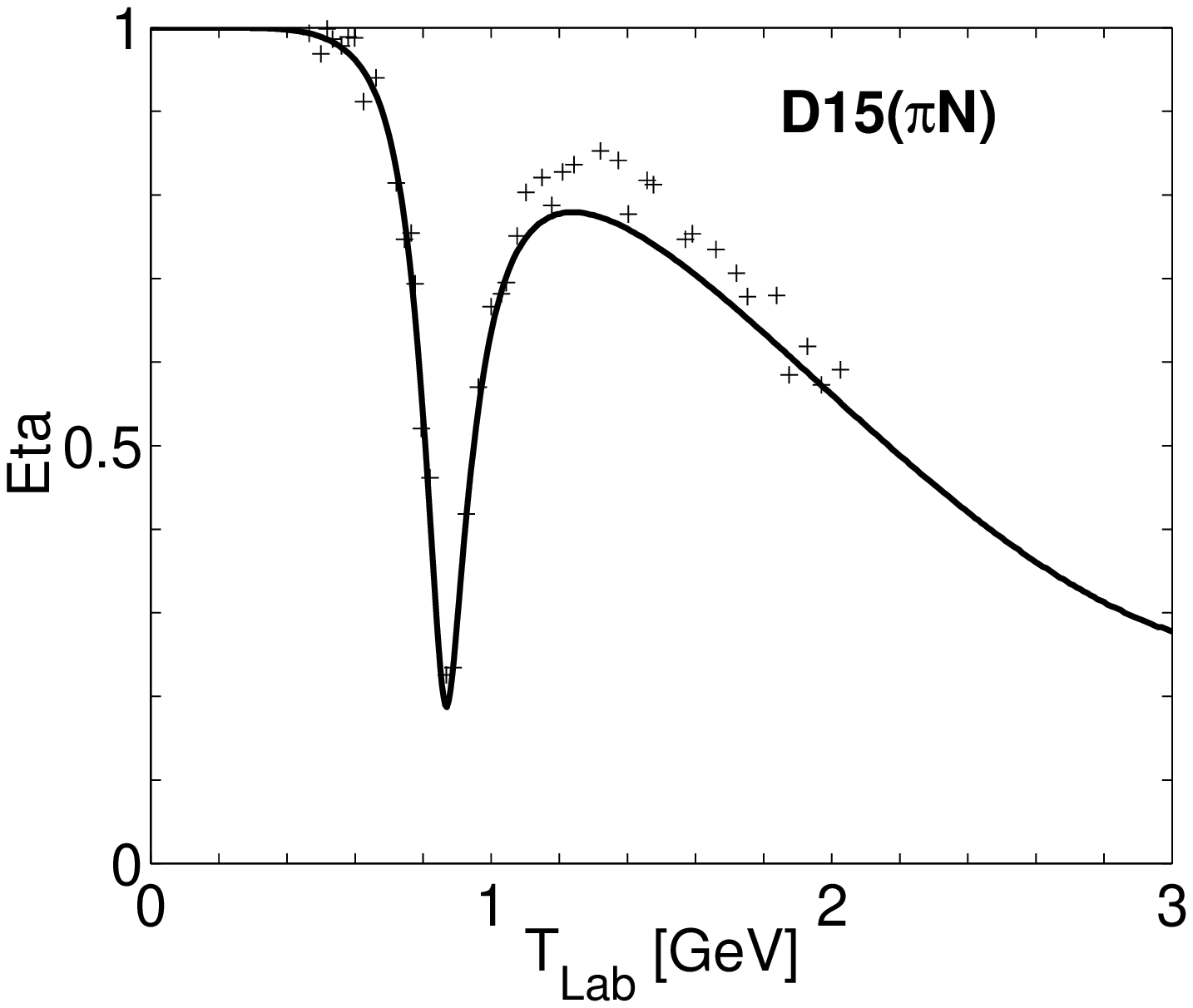,scale=0.37}
\end{figure}
\begin{figure}[ht]
\centering
\epsfig{file=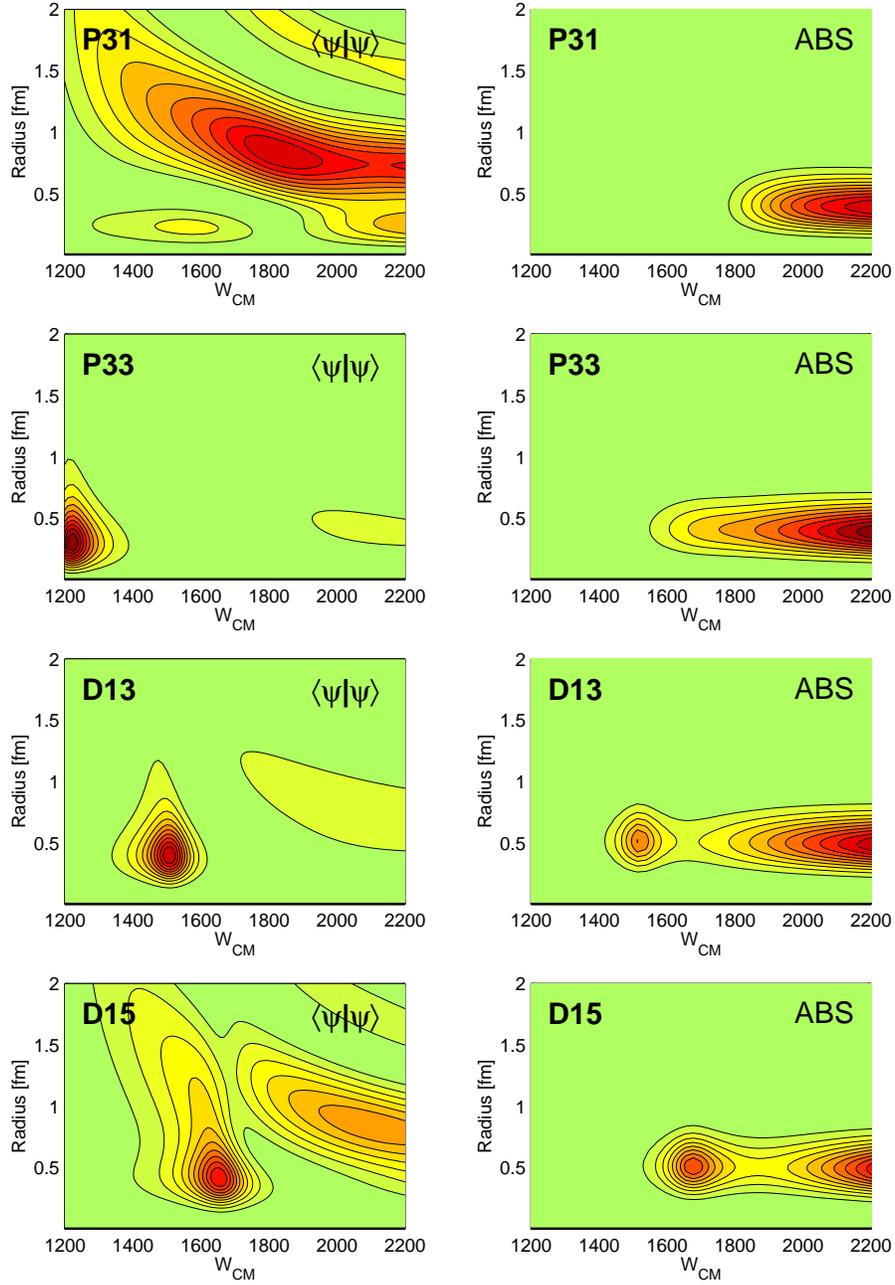,scale=0.7}
\caption{Sample of $\pi$N Phase shifts with  probabilities (left column) and current loss
(right column)}
\label{Fig1}
\end{figure}
\newpage
\begin{figure}[ht]
\centering
\epsfig{file=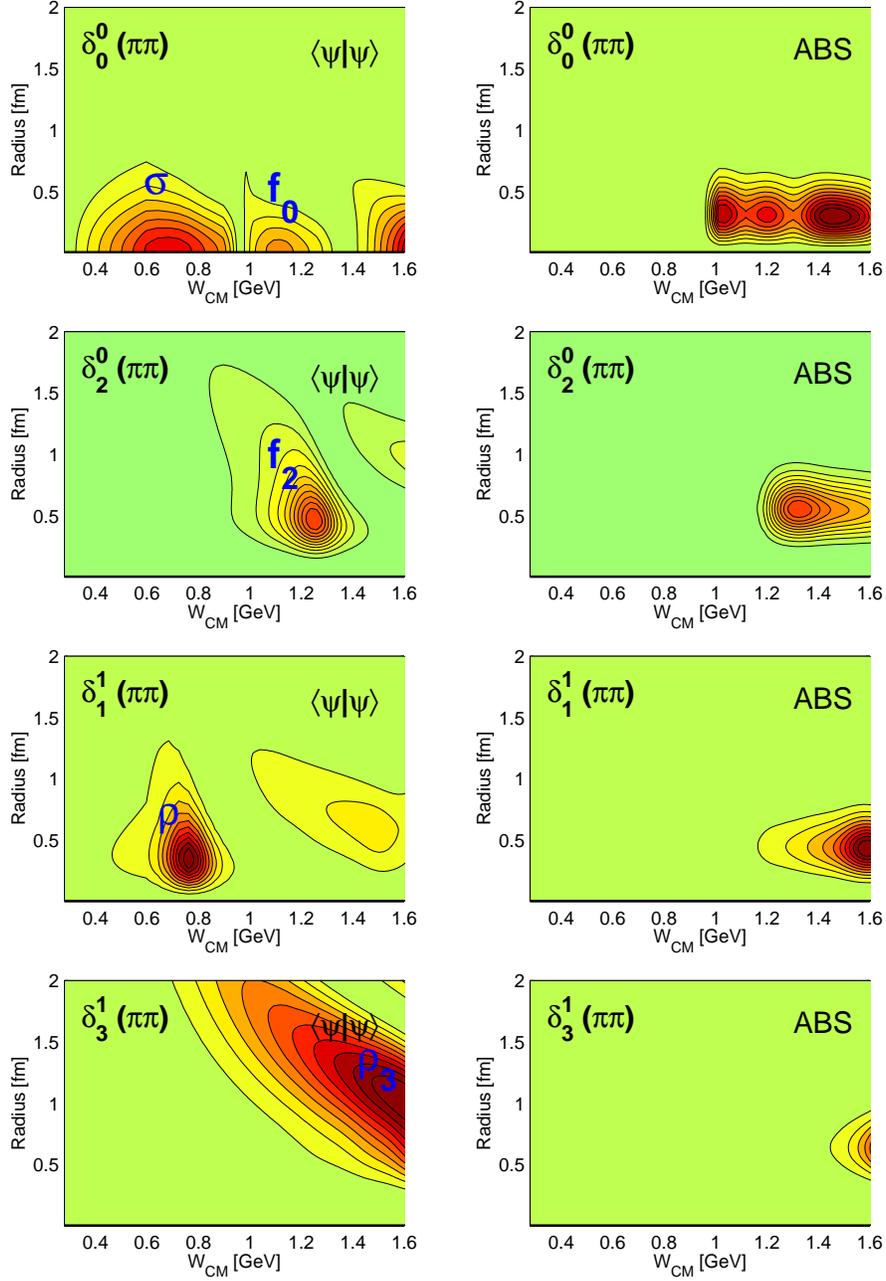,scale=0.7}
\caption{A sample of $\pi\pi$ probabilities and loss of currents. $\delta_1^1$ shows the
 $\rho$-meson and $\delta^0_2$ the $f_2$ resonance }
\label{Fig2}
\end{figure}
\end{document}